\documentclass[12pt]{article}
\usepackage [dvips]{graphicx} 

\newcommand{\be}{\begin{equation}}
\newcommand{\ee}{\end{equation}}
\newcommand{\ba}{\begin{array}}
\newcommand{\ea}{\end{array}}
\newcommand{\bc}{\begin{center}}
\newcommand{\ec}{\end{center}}
\newcommand{\disregard}[1]{{}}

\begin{document}

\title{Static Properties of Trapped Bose Gases at Finite Temperature:
Thomas-Fermi Limit}
\author{ M. Benarous and H. Chachou-Samet \\
{\it Laboratory for Theoretical Physics and Material Physics} \\
{\it Faculty of Sciences and Engineering Sciences} \\
{\it Hassiba Benbouali University of Chlef} \\
{\it B.P. 151, 02000 Chlef, Algeria.} }
\date{\today}
\maketitle

\begin{abstract}
We rely on a variational approach to derive a set of equations governing a
trapped self-interacting Bose gas at finite temperature. In this work, we
analyze the static situation both at zero and finite temperature in the
Thomas-Fermi limit. We derive simple analytic expressions for the condensate
properties at finite temperature. The noncondensate and anomalous density
profiles are also analyzed in terms of the condensate fraction. The results
are quite encouraging owing to the simplicity of the formalism.
\end{abstract}


\newpage

\setcounter{section}{0} \setcounter{equation}{0}

\begin{center}
{}
\end{center}

\section{Introduction}

In a remarkable series of experiments on Rubidium and Sodium vapors\cite{and95,dav95}, the Bose-Einstein condensation was first observed. Although
having been predicted theoretically a long time ago for noninteracting boson
systems\cite{BE24}, the experimental challenge was to demonstrate that a
real gas can indeed be ``bose condensed''. Since then, a great effort was
devoted by researchers all around the world in order to understand and
predict the condensate properties. The main tools, beside the Monte-Carlo
calculations\cite{KR}, were the Bogoliubov\cite{BG}, the Popov\cite{PO}, the
Beliaev\cite{BL,GR} and the Hartree-Fock-Bogoliubov\cite{HF,KT97,ST}
approximations. These approximations all adopt some simplifying assumptions
about the various quantities involved in the problem, such as the order
parameter $\Phi$, or the condensate density $n_c\equiv |\Phi|^2$, the
non-condensed density or thermal cloud $\tilde{n}$ and the anomalous density 
$\tilde{m}$. A major well-known drawback of these methods is that they
cannot be easily extended to situations where their main assumptions fail.
In a previous paper\cite{BM05}, we rely on a different approach, based on
the time-dependent variational principle of Balian and V\'en\'eroni\cite{BV}
, which allows one to overcome some of those restrictions. We obtained a set
of three coupled dynamical equations, which we called ``Time-Dependent
Hartree-Fock-Bogoliubov'' (TDHFB) equations, governing the evolution of $%
\Phi $, $\tilde{n}$ and $\tilde{m}$. They were shown to generalize in a
consistent way the Gross-Pitaevskii equation\cite{GP}.

The present paper is devoted to analyze the implications of our TDHFB
equations. Since it is important to apprehend first what happens in the
static situation before going further into the analysis of the excitation
spectrum or to the full dynamical case, we will focus on the static solution
both at zero and finite temperature in the Thomas-Fermi (TF) limit. The
interest is evident, since there remain many unanswered questions such as
the general dependence of the density profiles on the temperature and on the
interaction strength and the effect of the interactions on the critical
temperature. More particularly, recent experiments are raising challenging
questions about the precise determination of the thermal cloud and its
backeffects on the condensate\cite{gerbier}. Indeed, due to the difficulties
inherent to these experiments, there is no clear image on the way the
condensed and non condensed phases mix up. Hence, in these preliminary
calculations, we intend to provide some simple answers. We do not pretend of
course to reproduce exactly the experimental data or the full Monte-Carlo
calculations, but we would like to show that the simplifications that we are
actually using (Mean field + Thomas-Fermi) are controlable and retain also
the most important qualitative features without destroying the underlying
physics. This provides a simple enough tool which can be considered as a
starting point for a more elaborate treatment.

The paper is organized as follows. In section 2, we recall the main steps
that have been used in \cite{BM05} to derive the TDHFB equations. Then, we
present the static solutions and discuss their properties at zero
temperature. At finite temperature, the equations are much more involved and
require a careful analysis. In the TF limit, we present a simple method
which allows for a self-consistent determination of the various density
profiles as well as some other static properties of the condensate such as
the chemical potential and the condensate radius. Indeed, the TF
approximation obviously provides simple enough analytical expressions since
it neglects the kinetic terms thus yielding algebraic equations instead of
partial differential equations. This is the main advantage of our method
which yields the most important qualitative features without having to
handle highly non-linear differential equations.

In section 3, we present the results of our calculations. We plot first the
condensate radius and the central density as functions of the condensate
fraction and note in particular the compression effect of the condensate due
to the thermal cloud. Moreover, we discuss the TF profile obtained for the
condensate density even at low condensate fraction. The noncondensate
density profile is also plotted for a wide range of condensate fraction and
shows a good qualitative agreement with recent experiments. Finally, the
anomalous density, although not yet measured experimentally, is shown to
behave in a quite intuitive way.

Some concluding remarks are given at the end of the paper.
\setcounter{equation}{0}
\begin{center}
{}
\end{center}

\section{The TDHFB Equations and Their Static Solutions}

The general TDHFB equations were derived in ref.\cite{BM05} for a grand
canonical Hamiltonian of trapped bosons with quartic self-interactions (with
coupling constant $g$ and mass $m$): 
\begin{equation}
H=\int_{{\bf r}}a^{+}({\bf r})\left[ -{\frac{\hbar ^{2}}{2m}}\Delta +V_{{\rm 
{ext}}}({\bf r})-\mu \right] a({\bf r})+{\frac{g}{2}}\int_{{\bf r}}a^{+}(
{\bf r})a^{+}({\bf r})a({\bf r})a({\bf r}).  \label{eq1}
\end{equation}
The quantity $V_{{\rm {ext}}}({\bf r})$ is the trapping potential and $\mu $
is the chemical potential. These equations read: 
\begin{equation}
\begin{array}{rl}
i\hbar \dot{\Phi} & =\left( -{\frac{\hbar ^{2}}{2m}}\Delta +V_{{\rm {ext}}
}-\mu +gn_{c}+2g\tilde{n}\right) \Phi +g\tilde{m}\Phi ^{\ast }, \\ 
i\hbar \dot{\tilde{n}} & =g\left( \tilde{m}^{\ast }\Phi ^{2}-\tilde{m}{\Phi
^{\ast }}^{2}\right) , \\ 
i\hbar \dot{\tilde{m}} & =g(2\tilde{n}+1/V)\Phi ^{2}+4\left( -{\frac{\hbar
^{2}}{2m}}\Delta +V_{{\rm {ext}}}-\mu +2gn+{\frac{g}{4}}(2\tilde{n}
+1/V)\right) \tilde{m},
\end{array}
\label{eq2}
\end{equation}
where we have introduced the volume $V$ of the gas in order to ensure the
correct dimensions. In Eqs.(\ref{eq2}), $\Phi $ is the order parameter, $
n_{c}$ the condensate density ($n_{c}=|\Phi |^{2}$), $\tilde{n}$ the
non-condensed density (or thermal cloud) and $\tilde{m}$ is the anomalous
density. The quantity $n\equiv {n_{c}}+\tilde{n}$ is the total density.

The TDHFB equations with a general Hamiltonian $H$ were derived in \cite{BF99}. The properties discussed here and in \cite{BM05} were established
there in their general forms. These equations were obtained using the
Balian-V\'{e}n\'{e}roni varia\-tional principle\cite{BV}, with a gaussian
trial density operator (that is, an exponential operator of a quadratic
form) in the creation and annihilation operators. The result was a set of
coupled evolution equations for the expectation values 
$\langle a\rangle $, $\langle a^{+}a\rangle -\langle a^{+}\rangle \langle a\rangle $ and 
$\langle aa\rangle -|\langle a\rangle |^{2}$. When one identifies these quantities
respectively with the order parameter $\Phi $, the non-condensed density $\tilde{n}$ and the anomalous density $\tilde{m}$, and when one restricts $H$
to the class (\ref{eq1}), the equations (\ref{eq2}) follow.

The TDHFB equations couple in a consistent and closed way the three
densities. They should in principle yield the general time, space and
temperature dependence of the various densities. Furthermore, they obviously
constitute a natural extension of the Gross-Pitaevskii equation\cite{GP}.
They are not only energy and number conserving, but also satisfy the
Hugenholtz-Pines theorem (see below) which leads to a gapless excitation
spectrum in the uniform limit. Moreover, the two last equations in (\ref{eq2}) are not totally independent since $\tilde{n}$ and $\tilde{m}$ are related
by the \textquotedblright unitarity\textquotedblright\ relation\cite{BM05}:
\begin{equation}
I=\left( 1+2V\tilde{n}\right) ^{2}-\left( 2V|\tilde{m}|\right) ^{2},
\label{eq2t}
\end{equation}
where the Heisenberg parameter $I$ (which is always $\geq 1$) is a measure
of the temperature, the lower limit being the zero temperature case. For
instance, for a thermal distribution at equilibrium, $I$ writes as $I=\coth
^{2}{(\hbar \omega _{0}/2{k_{B}}T)}$, where $\omega _{0}$ is the average
frequency of the trapping field\cite{BM05} 
\footnote{In fact, one can show that for a system of energy $E$, $\sqrt{I}
=1+2\,f_{B}(E)$, where $f_{B}$ is the Bose-Einstein distribution.}
. We therefore see that upon replacing $\tilde{n}$ by its expression given in (\ref{eq2t}), the temperature appears explicitly in the equations.

It is to be mentioned that the TDHFB equations have also been derived by
several authors using different variational formulations\cite{CHER03,Pr04}.
In the first reference, the authors have obtained a set of equations very
similar to ours. In fact, we can show that our equations can be deduced from
theirs by taking the diagonal elements (${\bf r}={\bf r^{^{\prime }}}$) of
the equations (B1), (B2) and (B3) 
\footnote{There is however a factor $1/2$ missing from (B3).} of ref.\cite{CHER03}.

The static solutions, which are the object of our study in this work, are
obtained by setting to zero the right hand sides of (\ref{eq2}). At zero
temperature, the standard TF limit\cite{SV} amounts to neglecting the kinetic
(or $\Delta $) term in the Gross-Pitaevskii equation. This is particularly
satisfied for strong interacting regimes or large atom numbers. At finite
temperature and below the transition, since there are two phases (condensed
and non condensed) which coexist, one has to provide a complementary recipe
for what we shall call the finite temperature TF limit. First, neglecting
the kinetic energy of the condensate remains a justifiable approximation
since the atoms are slowed down in order to obtain condensation. On the
other hand, $\tilde{m}$ is believed to be an extremely small and slowly
varying function whatever the temperature is (recall that it describes the
correlations between the condensed and non-condensed phases). Hence, one may
in a first approximation safely neglect $\Delta \tilde{m}$. Heuristically, one may argue
that, since the equations for $n_c$ and $\tilde{m}$ contain almost comparable operators,
$h_0$ and $h_0+g(n_c + (1+2V\tilde {n})/4V)$, where $h_0$ is the self-consistent mean field hamiltonian 
$h_0= V_{{\rm {ext}}}(r)-\mu +g{n_{c}}+2g\tilde{n}$, the TF condition $h_0 >> T$ ($T$ being the kinetic 
operator), if fullfilled for $n_c$ shouldf also be satisfied for $\tilde{m}$. For this approximation to be consistent, 
$n_c$ and $\tilde{m}$ should vary on the same characteristic length, which is indeed the case as we will show later.

Before proceeding further, it is important to notice at this point that a kinetic-like term of the thermal cloud does
not appear explicitely in the equations but is rather hidden in the third
equation of (\ref{eq2}). Indeed, the kinetic term of the thermal cloud is
related to the second derivative of the anomalous density. Differentiating (
\ref{eq2t}) yields a relation of the form:
\begin{equation}
\Delta \tilde{n}\sim \left( \nabla |\tilde{m}|\right) ^{2}-\left( \nabla |
\tilde{n}|\right) ^{2}+|\tilde{m}|\Delta |\tilde{m}|,  \label{eq2r}
\end{equation}
which shows in particular that neglecting $\Delta \tilde{m}$ does not
necessarily mean neglecting $\Delta \tilde{n}$. That is precisely the
recipe that we shall adopt below.

With this finite temperature prescription, the static equations
corresponding to (\ref{eq2}) now write
\begin{equation}
\begin{array}{rl}
& \left( V_{{\rm {ext}}}(r)-\mu +g{n_{c}}+2g\tilde{n}\right) \Phi +g\tilde{m}
\Phi ^{\ast }=0, \\ 
& \tilde{m}^{\ast }\Phi ^{2}-\tilde{m}{\Phi ^{\ast }}^{2}=0, \\ 
& \left( V_{{\rm {ext}}}(r)-\mu +2gn+{\frac{g}{4V}}(2V\tilde{n}+1)\right) 
\tilde{m}+{\frac{g}{4V}}(2V\tilde{n}+1)\Phi ^{2}=0,
\end{array}
\label{eq2s}
\end{equation}
These equations are naturally gapless and satisfy the Hugenholtz-Pines
theorem\cite{HF}. Indeed, owing to the second equation in (\ref{eq2s}), one
can easily show that at zero momentum, the relation 
$\mu =g(n+\tilde{n}-|\tilde{m}|)$ is clearly satisfied without adding further assumptions, as is
usually performed\cite{HF}.

In order to solve these equations, we may distinguish two rather different
situations. The first one is for $T=0$. When all the atoms are condensed, 
$\tilde{n}=\tilde{m}=0$, and ${{n_{c}}}$ equals the total density $n$ of the
gas. Omitting the trivial solution with ${{n_{c}}}=0$, one may take into
account just the first equation in (\ref{eq2s}), since we consider a gas
without a quantum cloud. Indeed, within the present set of equations, it is
an approximation (although justifiable) to ignore the quantum depletion at $T=0$. 
The last two equations in (\ref{eq2s}) become therefore meaningless,
and we are left with a simple expression for the condensate density 
\begin{equation}
{{n_{c}}}(r)=-\xi (r)={\frac{1}{g}}\left( \mu -V_{{\rm {ext}}}(r)\right) .
\label{eq4}
\end{equation}
Upon defining the oscillator length $a_{0}=(\hbar /m\omega _{0})^{1/2}$ and
the s-wave scattering length $a=mg/4\pi \hbar ^{2}$, we obtain for a
spherical trapping potential $V_{{\rm {ext}}}(r)={\frac{1}{2}}m\omega
_{0}^{2}r^{2}$, the condensate radius $R$ and the reduced chemical potential 
$\nu _{0}=\mu /{{\frac{1}{2}}}\hbar \omega _{0}$ for a gas of $N$ bosons as
\begin{equation}
\frac{R}{a_{0}}=\left( 15N\frac{a}{a_{0}}\right) ^{1/5},  \label{eq5}
\end{equation}
\begin{equation}
\nu _{0}=\left( 15N\frac{a}{a_{0}}\right) ^{2/5}.  \label{eq55}
\end{equation}
The preceding expressions show that the spreading of the condensate depends
essentially on the balance between the self-interactions and the trapping
potential. These results have also been obtained by many other authors, see
e.g. \cite{GR,ST,BM05}.

When $0\leq T<T_{{\rm {BEC}}}$, we have of course ${{n_{c}}}\neq 0$ and $
\tilde{n}\neq 0$. Let us introduce the parametrization\ $2V\tilde{n}+1=\sqrt{
I}\cosh {\sigma }$, $2V|\tilde{m}|=\sqrt{I}\sinh {\sigma }$, which
automatically endows the relation (\ref{eq2t}). Then, from the third
equation in (\ref{eq2s}), one obtains a simple equation for $X=e^{\sigma }$:
\begin{equation}
3X^{4}-4X^{2}+1+\frac{4Vn_{c}}{\sqrt{I}}\left( X^{2}-3\right) X=0,
\label{eq7}
\end{equation}
from which one extracts $\tilde{n}$ and $|\tilde{m}|$ as functions of $n
_c$. Next, one uses these expressions in the first equation (\ref{eq2s})
to get the condensate density
\begin{equation}
{{n_{c}}}(r)=-\xi (r)-{\frac{1}{V}}\left( \frac{X+3X^{-1}}{4}\sqrt{I}
-1\right) .  \label{eq66}
\end{equation}
What is remarkable is that the sole acceptable solution of equation (\ref
{eq7}) is a bounded function of $\eta =Vn_{c}/\sqrt{I}$. It is represented
on figure 1.  

\begin{figure}[h]
\includegraphics[scale=.7]{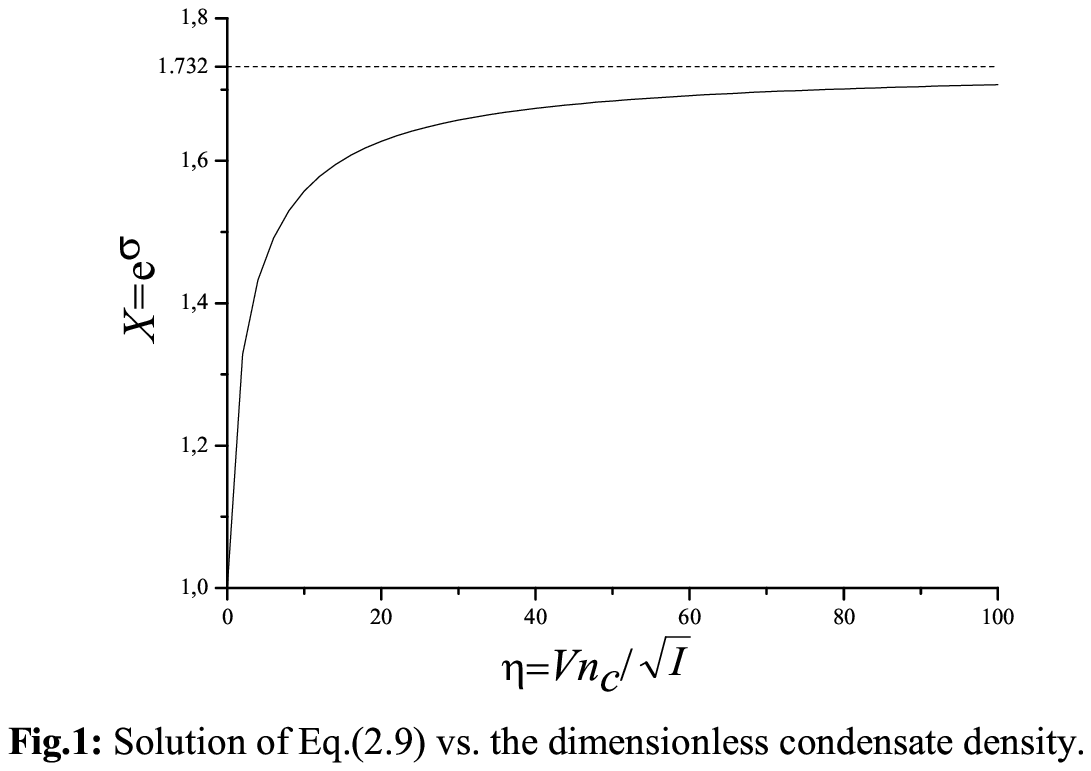}
\hfill
\includegraphics[scale=.7]{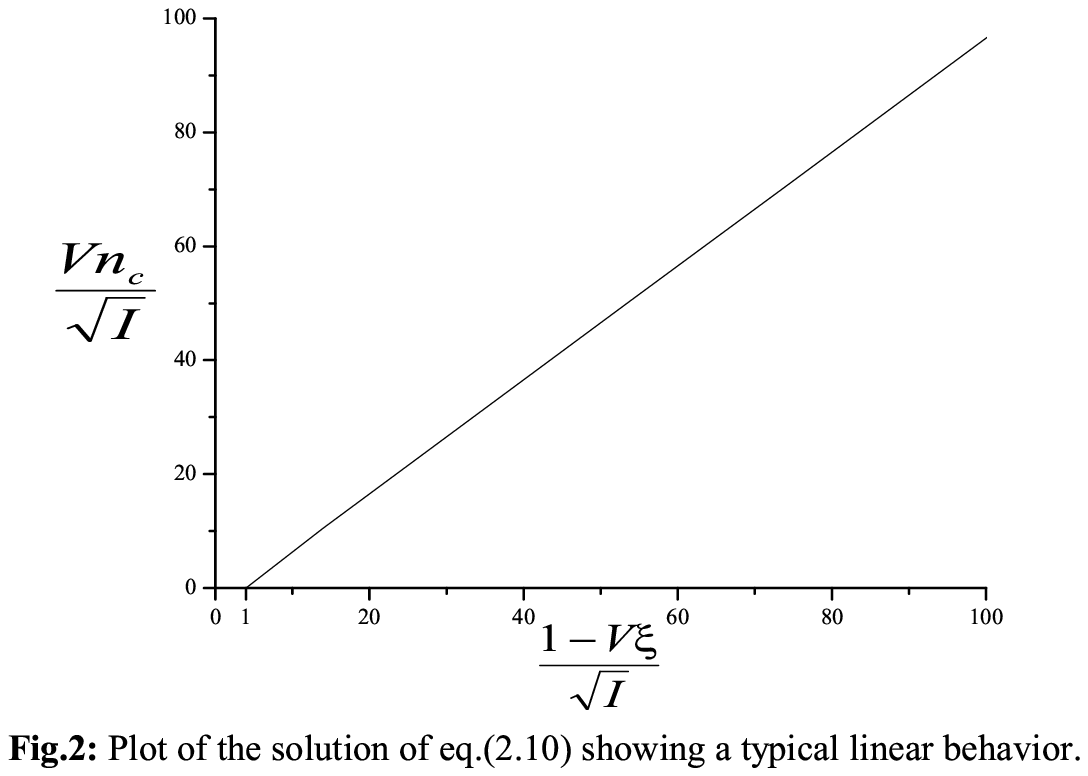}
\end{figure}

Due to this behavior, one can easily show that the quantity $\frac{X+3X^{-1}}{4}$ 
which appears in (\ref{eq66}) is almost independent of $n_{c}$ and becomes rapidly 
close to unity. Indeed, since equation (\ref{eq66}) 
may also be rewritten as $\eta ={\frac{(1-V\xi )}{\sqrt{I}}}-(X+3X^{-1})/4$, 
its solution provides the typical linear behavior shown in figure 2.

Hence, one may safely approximate (\ref{eq66}) by
\begin{equation}
{{n_{c}}}(r)\simeq -\xi (r)-{\frac{1}{V}}\left( \sqrt{I}-1\right) .
\label{eq67}
\end{equation}
In fact, one can check that the relative error between the two expressions (
\ref{eq66}) and (\ref{eq67}) is less than 1\%. Finally, since $\sqrt{I}$
does not depend on space, the result (\ref{eq67}) shows that the finite
temperature correction to the Thomas-Fermi profile (\ref{eq4}) is simply a
space-independent (but temperature dependent) shift. This shift may be
absorbed in a redefinition of the chemical potential which now writes
\begin{equation}
{\mu }=V_{{\rm {ext}}}(R)+{\frac{g}{V}}\left( \sqrt{I}-1\right) ,
\label{eq667}
\end{equation}
where $R$ is the condensate radius. The condensate density finally writes in
the suggestive form
\begin{equation}
{{n_{c}}}(r)=\frac{V_{{\rm {ext}}}(R)-V_{{\rm {ext}}}(r)}{g},  \label{eq8}
\end{equation}
which is formally the zero temperature TF profile. It is then easy to show
that the condensate radius takes also a simple form
\begin{equation}
\frac{R}{a_{0}}=\left( 15N_{c}\frac{a}{a_{0}}\right) ^{1/5},  \label{eq9}
\end{equation}
but now, it is the number of condensed atoms $N_{c}$ which is involved and
not the total number of atoms. The same conclusion may be drawn for the
chemical potential (\ref{eq667}). Hence, our finite temperature prescription
for the TF approximation provides natural extensions of the zero temperature
expressions, since the Thomas-Fermi parameter is now $N_{c}a/a_{0}$ instead
of $Na/a_{0}$.

In order to apprehend better these results, let us compute the remaining
unkown quantities, such as the non condensed and the anomalous densities. To
this end, and in order to obtain tractable expressions, we find it more
convenient to use the simple fit
\begin{equation}
X={\frac{\sqrt{3}\eta +{2/3}}{\eta +{2/3}},}  \label{eq68}
\end{equation}
(instead of the full analytical solution of equation (\ref{eq7})) which
reproduces correctly the solution $X$ plotted in\ figure 1 with a residual
error less than 0.1\%. Upon rewriting equation (\ref{eq8}) in the form $\eta
=\eta _{0}(1-x^{2})$, with an obvious definition of $\eta _{0}$, we obtain
the non condensate density
\begin{equation}
\tilde{n}(x)=\frac{1}{2V}\left\{ \frac{\sqrt{I}}{2}\left( \frac{\sqrt{3}\eta
_{0}(1-x^{2})+{2/3}}{\eta _{0}(1-x^{2})+{2/3}}+\frac{\eta _{0}(1-x^{2})+{2/3}
}{\sqrt{3}\eta _{0}(1-x^{2})+{2/3}}\right) -1\right\} {,}  \label{eq10}
\end{equation}
and the anomalous density
\begin{equation}
\left\vert \tilde{m}\right\vert (x)=\frac{1}{2V}\frac{\sqrt{I}}{2}\left( 
\frac{\sqrt{3}\eta _{0}(1-x^{2})+{2/3}}{\eta _{0}(1-x^{2})+{2/3}}-\frac{\eta
_{0}(1-x^{2})+{2/3}}{\sqrt{3}\eta _{0}(1-x^{2})+{2/3}}\right) {,}
\label{eq11}
\end{equation}
as functions of the reduced radial distance $x=r/R$. In figures 3 and 4, we
show typical density profiles (in reduced units) for $\eta _{0}=1$. 

\begin{figure}[h] 
\includegraphics[scale=.7]{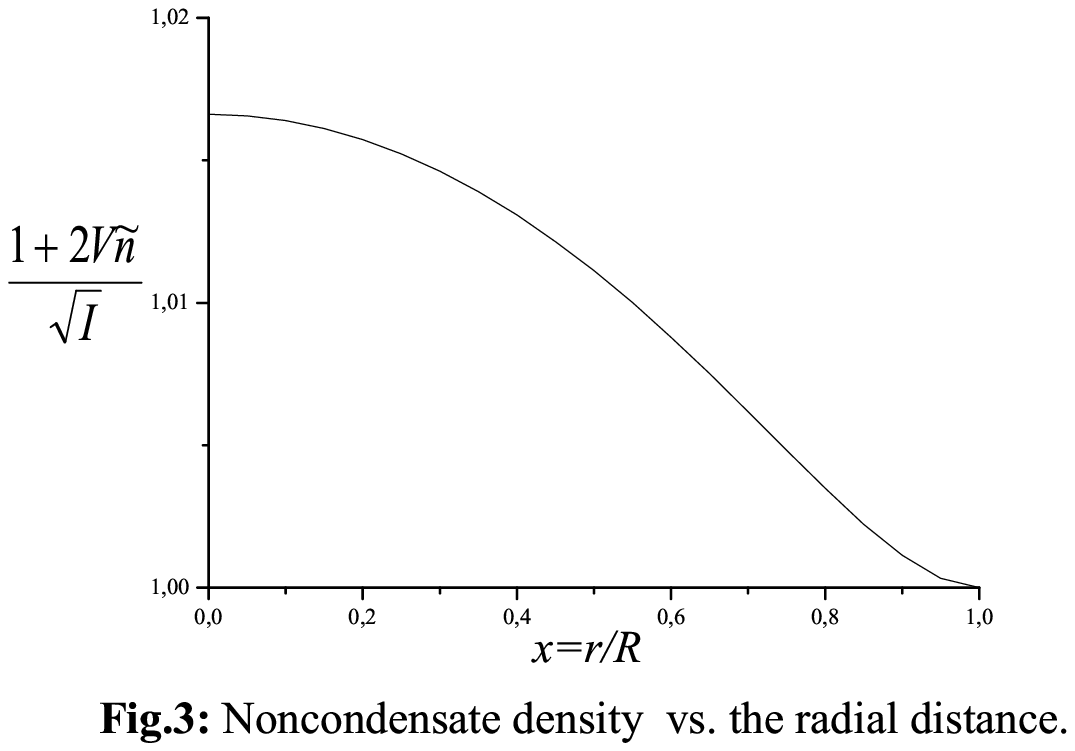} 
\hfill
\includegraphics[scale=.7]{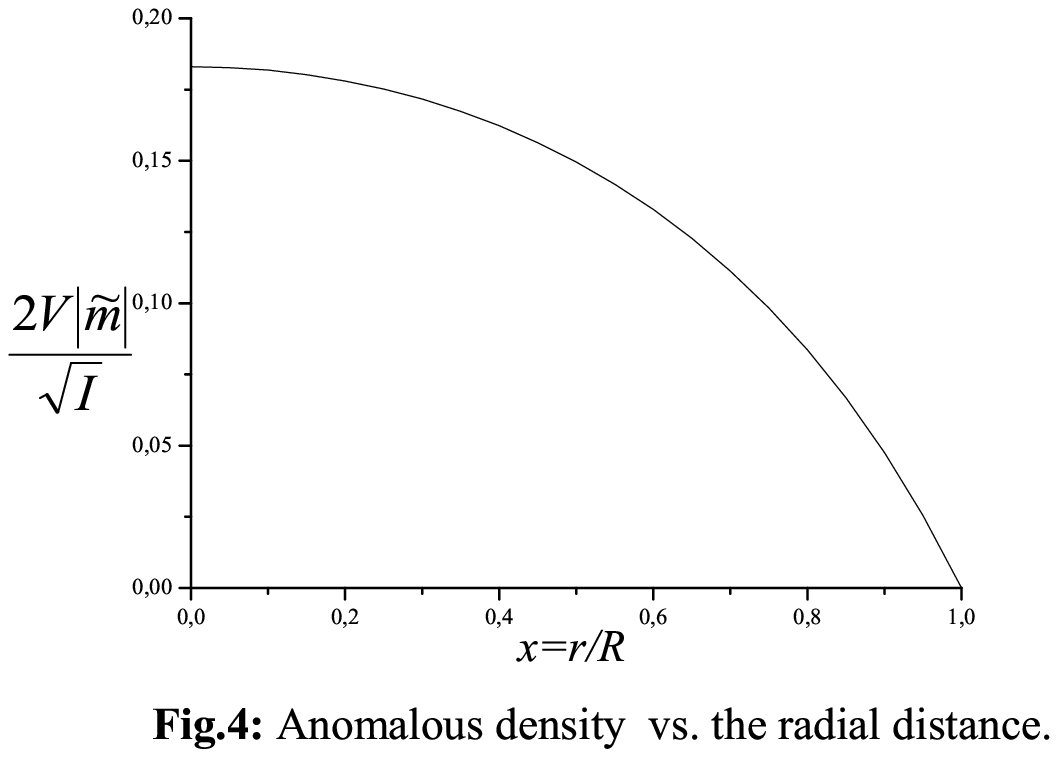} 
\end{figure} 

It is interesting to notice that, within our TF approximation, the figure 3
is qualitatively consistent with the profile of the thermal cloud as depicted
by Gerbier {\it et al}\cite{gerbier}. It also predicts that the thermal
cloud does not vanish at the boundaries of the condensate which is also
compatible with the experimental results since it is widely known that
condensed atoms are surrounded by the thermal cloud. What is less known
(even experimentally) is the anomalous density (figure 4). This quantity behaves 
quite differently from the thermal cloud and is clearly dominated by a TF-like shape. 
In contrast with the thermal cloud, the anomalous density vanishes at the boundaries
which is also plausible since the condensate vanishes there. Furthermore, we observe that 
$n_c$, $|\tilde{m}|$ and $\tilde{n}$ vary on the same length scale ($R$) which justifies 
{\it a posteriori} our previous assumption.

In order to obtain more quantitative results, one must determine $N_{c}$ by
using the normalization condition. We get easily the relation
\begin{equation}
1+2N=2N_{c}+\sqrt{I}g(s),  \label{eq12}
\end{equation}
where
\begin{equation}
g(s)=\frac{2}{\sqrt{3}}+(\sqrt{3}-1)s\left\{ 1-\frac{3}{2}\sqrt{s+1}{\rm arc}\tanh \frac{1}{\sqrt{s+1}}
+\frac{1}{2}\sqrt{\frac{s}{\sqrt{3}}+1}{\rm arc}\tanh \frac{1}{\sqrt{\frac{s}{\sqrt{3}}+1}}\right\} ,  \label{eq13}
\end{equation}
with $s=4\sqrt{I}/15N_{c}$. But since the function $g(s)$ satisfies $1\leq
g(s)\leq 2/\sqrt{3}$, the equation (\ref{eq12}) is approximately solved to
yield, to a very good accuracy, the simple result
\begin{equation}
N_{c}\simeq N-\frac{\sqrt{I}-1}{2}.  \label{eq144}
\end{equation}
All the unknown quantities may now be determined in terms of $N$ and $\sqrt{I}$ alone. 
The corresponding results will be discussed in the next section.

\setcounter{equation}{0}

\section{Results and Discussions}

First of all, the condensate radius (\ref{eq9}) may be written as
\begin{equation}
R=R_{{\rm {TF}}}\left( \frac{N_{c}}{N}\right) ^{1/5},  \label{eq21}
\end{equation}
where $R_{{\rm {TF}}}$ is the zero temperature result given by equation (\ref
{eq5}). Figure 5 represents the condensate radius (in units of $R_{{\rm {TF}}
}$) as a function of the condensate fraction and we notice in particular the
compression of the condensate when reducing $N_{c}/N$ (that is increasing
the temperature). This effect is by now a well established experimental
result \cite{gerbier} and is attributed to the thermal cloud. The same
effect of compression is observed on figure 6 for the central condensate
density $n_{c}(r=0)$ but it is more pronounced due to the power law of $2/5$
(see \ref{eq8}) instead of $1/5$ for the condensate radius. To be more
precise, let us choose generic values for the number of atoms and the
interaction strength ($N=10^{5}$ and $a/a_{0}=0.5\,10^{-3}$) and plot the
various densities (in units of the oscillator volume $a_{0}^{3}$) versus the
radial distance (in units of$\ R_{{\rm {TF}}}=3.758a_{0}$) for a condensate
fraction ranging from $5\%$ up to $60\%$.

\begin{figure}[h] 
\includegraphics[scale=.7]{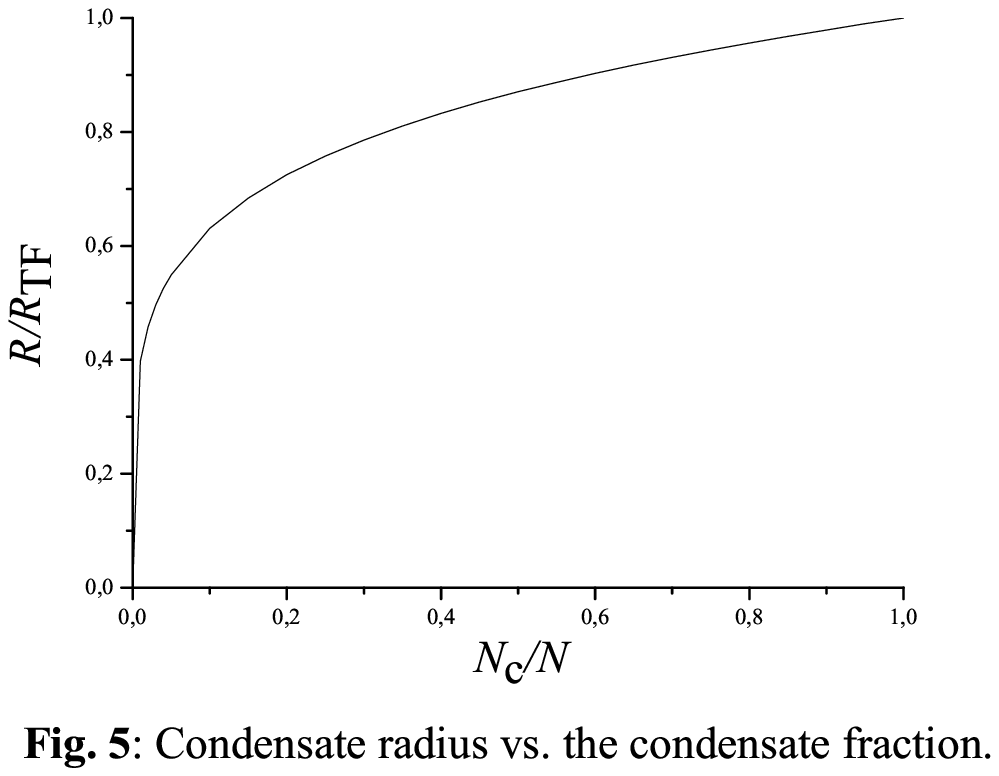} 
\hfill
\includegraphics[scale=.7]{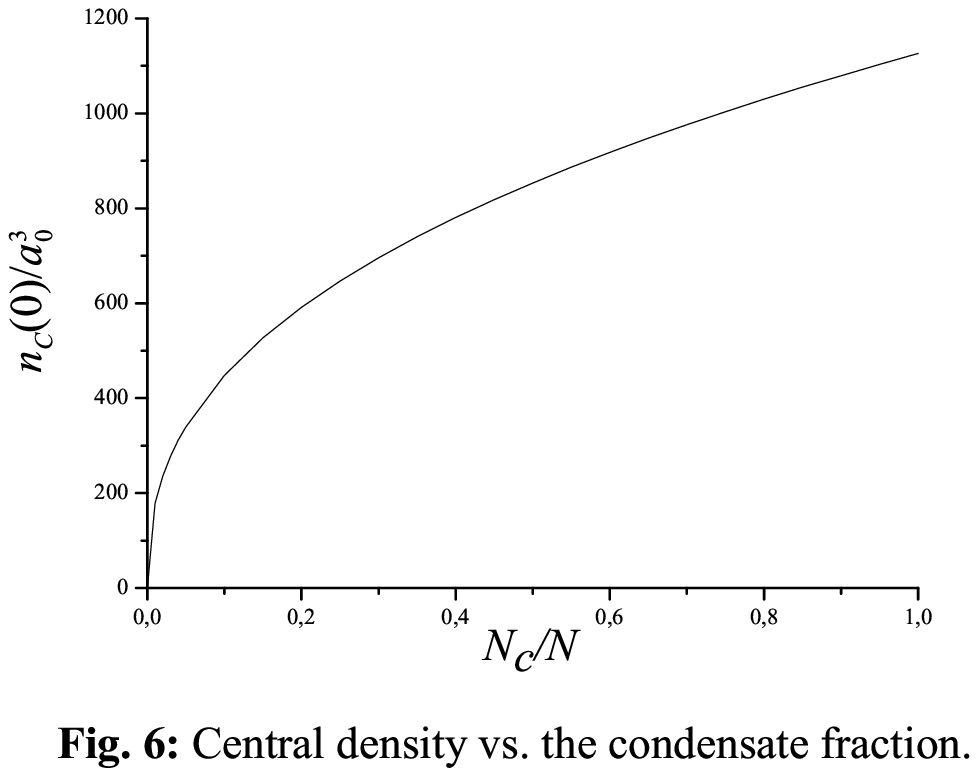} 
\end{figure}

The figure 7 shows typical Thomas-Fermi profiles for the condensate density,
even for low condensate fraction. This is of course what one may expect on
general grounds in the TF regime. Moreover, the effect of compression of the
condensate is also clearly visible here.

\begin{figure}[h] 
\begin{center} 
\includegraphics[scale=.8]{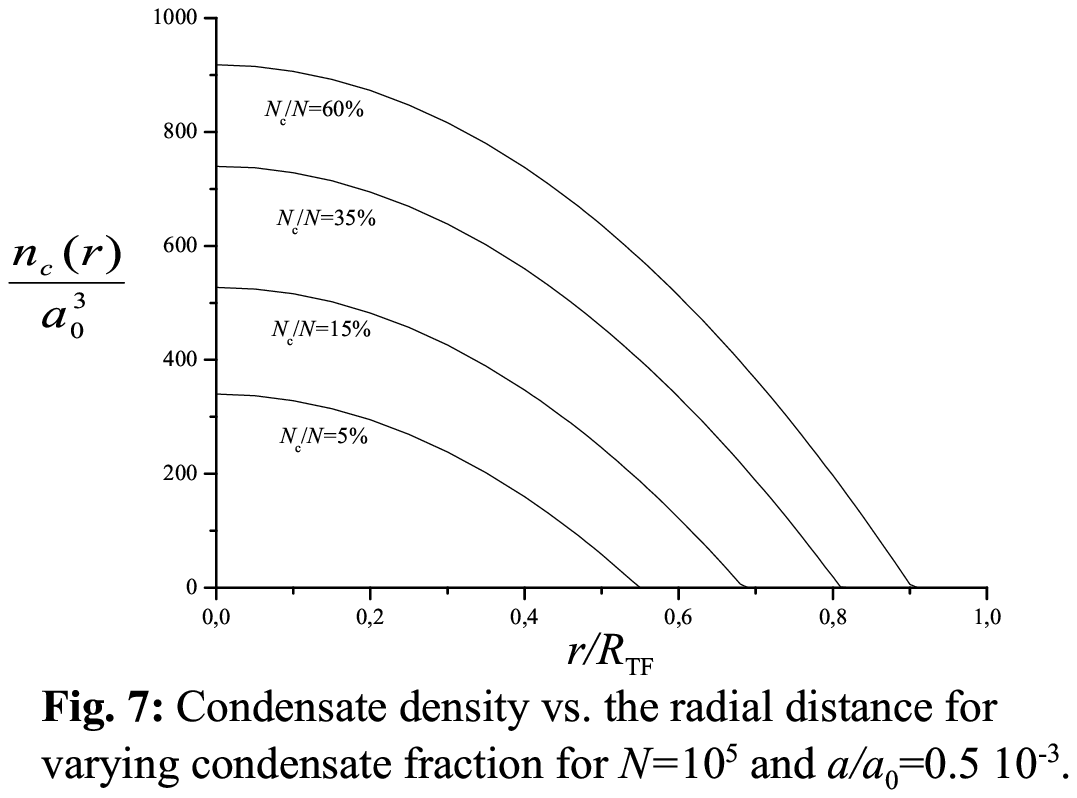} 
\end{center} 
\end{figure} 

The noncondensate density (from which we have substracted a constant $\tilde{
n}(R)$ for clarity) is plotted on figure 8 with the same units as before.
\begin{figure}[h] 
\begin{center} 
\includegraphics[scale=.8]{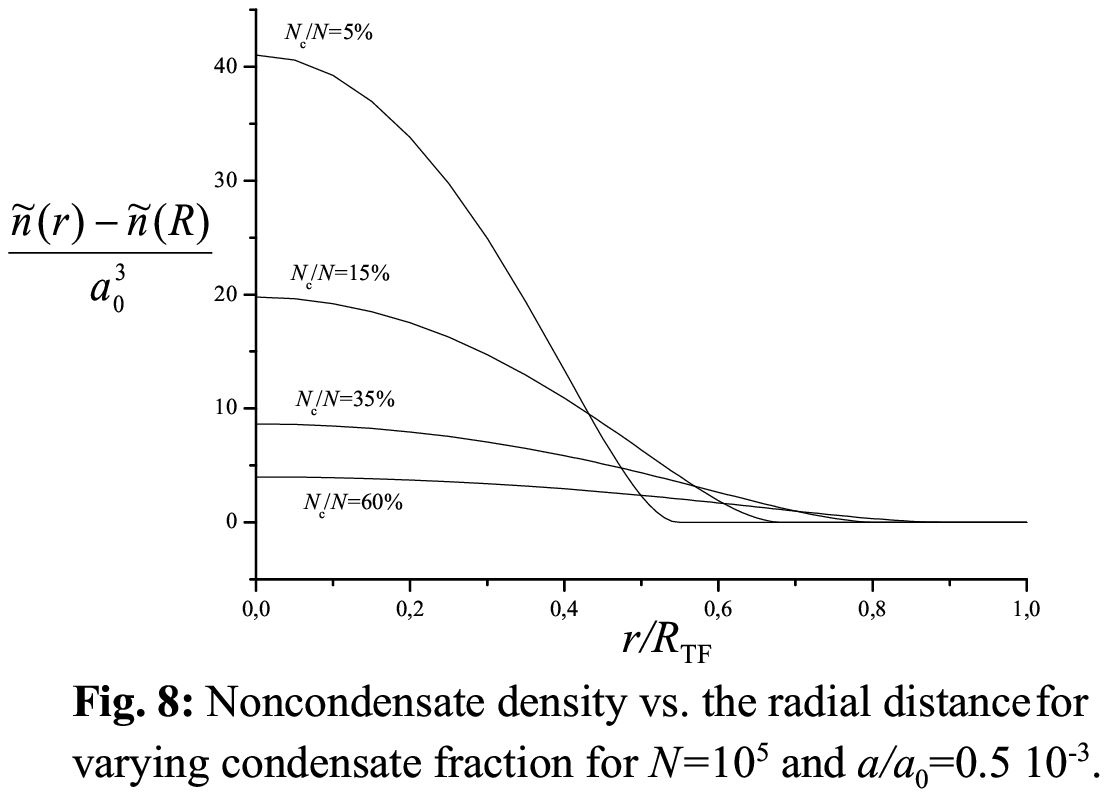} 
\end{center} 
\end{figure} 
The information which is added here with respect to figure 3 is the
temperature dependence which appears via the condensate fraction. As noted
earlier, the experimental result\cite{gerbier} is well reproduced
qualitatively for the whole range of condensate fraction\footnote{
Although, in order to compare exactly with the experimental results, one
must include an overall scale factor due to the finite ballistic expansion
time.}. In particular, we notice that when increasing the condensate
fraction, the thermal cloud tends to spread and flatten. On the other hand,
the thermal cloud takes on a (small but) finite value for $r\geq R$. Even if
this behavior is less intuitive, it is not very surprising since we do know
that neglecting the second order derivatives amounts to making a cut of the
densities at the boundaries. It is indeed a limitation of the TF
approximation at the boundaries\cite{dalfovo}. The tail should be reproduced when one
reinjects the second derivatives in the equations.

\begin{figure}[h] 
\begin{center} 
\includegraphics[scale=.8]{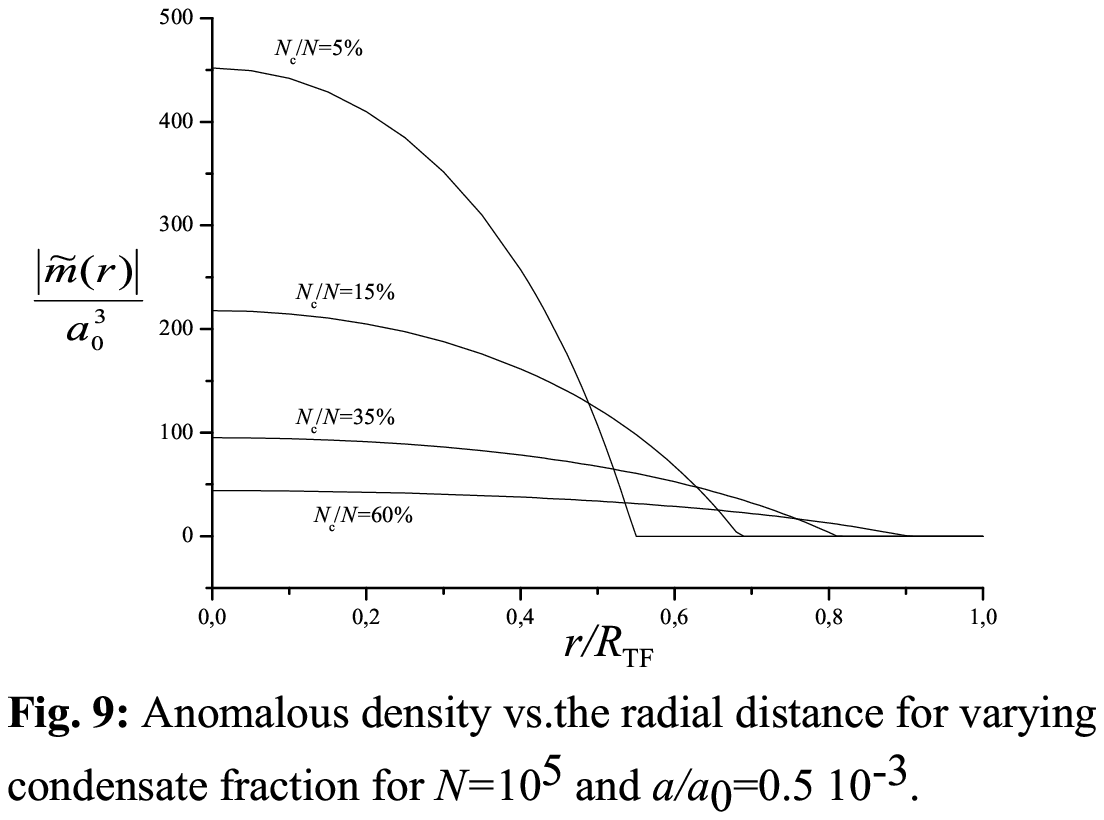} 
\end{center} 
\end{figure} 

Finally, we plot on figure 9 the anomalous density. To our knoweledge, this
quantity was never measured experimentally,\ and it is interesting that our
calculations predict a very simple and yet intuitive behavior as well which
remains of course to be confirmed.

\setcounter{equation}{0}{}

\section{Concluding Remarks}

We present in this paper a finite temperature analysis of the static TDHFB
equations (derived in a previous paper) in the Thomas-Fermi limit for a gas
of bosons in a harmonic trap.

At zero temperature, we obtain familiar expressions for the chemical
potential and the condensate radius. The standard Thomas-Fermi profile for
the condensate density is also recovered.

At finite temperatures and below the transition, since there are two phases,
one should provide a prescription for the TF limit. We propose such a recipe
(maybe the simplest) which consists in neglecting the second order
derivatives of the condensate density and the anomalous density. The
underlying idea is that, although the anomalous density is necessary for the
coherence of the equations, it is believed to be a very small and hopefully
a very smooth quantity. We therefore obtain analytical expressions for the
condensate density, the condensate radius, the chemical potential and the
condensate fraction as functions of the temperature. Our expressions appear
as natural extensions of the zero temperature TF limit, since the relevant
parameter which controls the approximation becomes $N_{c}a/a_{0}$ instead of 
$Na/a_{0}$.

Most importantly, we derive quite simple expressions for the noncondensate
density and for the anomalous density, which we plot as functions of the
condensate fraction and draw many conclusions. First of all, the condensate profile is
almost of the TF shape of which the spatial extension and the heights are
controlled by $N_{c}/N$. Furthermore, the compression of the condensate by
the thermal cloud with increasing temperature is clearly visible. On the
other hand, the noncondensate density profile is qualitatively consistent with
the condensate fraction dependence observed in recent experiments. In
particular, the thermal cloud tends to spread and flatten with increasing
temperature. The calculated anomalous density, although not yet observed
experimentally, shows also a very intuitive behavior; it is maximal at the
center of the trap and zero at the boundaries. The tendency to spreading and
flattening with increading temperature is also observed here.

At the borders of the trap (where the condensate density vanishes) and for a
given temperature, the non condensate density takes on a finite value which
is a quite abrupt behavior. Athough this meets the fact that the thermal
cloud is actually surrounding the condensate, it is to by no means conclusive. 
But this is also a shortocoming of the TF approximation as a whole
since it is known to break down at the boundaries of the condensate. Indeed,
reinjecting the second derivatives of the densities will certainly entail a more
physical behavior.

We are grateful to P. Schuck and Y. Castin for fruitful discussions and a
careful reading of the manuscript.

One of us (M. B.) is thankful to the members of the Groupe de Physique 
Th\'{e}orique, IPN-Orsay-France, where part of this work has been done.

\end{document}